\documentclass{article}
\usepackage[portuguese]{babel}
\usepackage{graphicx}
\usepackage{epsfig}
\begin{document}

\begin{titlepage}
\title{Teorias Neo-Newtonianas}

\author{J\'ulio~C.~Fabris\footnote{fabris@pq.cnpq.br}\, e Hermano E.S. Velten\footnote{velten@pq.cnpq.br}\vspace{0.5cm}\\
Departamento de F\'{\i}sica - UFES, Vit\'oria, ES, Brasil \\
}
\date{30 de novembro de 2014}

\maketitle

\begin{abstract}

A teoria da Relatividade Geral \'e a moderna teoria da gravita\c{c}\~ao, tendo substituido a teoria newtoniana na descri\c{c}\~ao dos fen\^omenos
gravitacionais. No entanto, apesar dos grandes sucessos obtidos pela teoria da Relatividade Geral, a teoria gravitacional newtoniana continua
sendo largamente empregada devido ao fato que a teoria da Relatividade Geral incorpora, na maior parte dos casos, apenas pequenas corre\c{c}\~oes \`as predi\c{c}\~oes newtonianas. Al\'em disto, a teoria newtoniana possui uma grande simplicidade t\'ecnica e conceitual quando comparada com a teoria relativista.
Neste texto, discutimos a possibilidade de estender a teoria newtoniana tradicional de forma a incorporar efeitos tipicamente relativistas mas guardando a referida simplicidade t\'ecnica e conceitual. Denominamos estas extens\~oes de {\it teorias neo-newtonianas}. Estas teorias s\~ao discutidas principalmente nos contextos cosmol\'ogico e da astrof\'{\i}sica de objetos compactos.
\end{abstract} 
\end{titlepage}

\section{Introdu\c{c}\~ao}

A teoria newtoniana da gravita\c{c}\~ao, expressa matematicamente pela lei do inverso do quadrado da dist\^ancia, foi durante muito tempo um dos sucessos
cient\'{\i}ficos mais not\'aveis da hist\'oria da ci\^encia. Durante pelo menos os dois primeiros s\'eculos que seguiram sua formula\c{c}\~ao, a gravita\c{c}\~ao newtoniana explicou todos os fen\^omenos gravitacionais observados, desde a queda dos corpos na superf\'{\i}cie terrestre at\'e o movimento dos planetas
no sistema solar. Quando se acreditou que ela estava errada, pois Urano parecia seguir uma \'orbita distinta da prevista, a certeza que ela era uma teoria correta levou \`a predi\c{c}\~ao da exist\^encia de um outro planeta, com massa e \'orbita bem definidas. A descoberta de Netuno, com todas as caracter\'{\i}sticas previstas, representou um triunfo indubit\'avel: n\~ao apenas a teoria explicava o que j\'a se conhecia, mas tinha poder preditivo, sugerindo a exist\^encia do que ainda n\~ao havia sido observado. Mesmo hoje, quando a teoria newtoniana n\~ao \'e tida mais como a correta teoria gravitacional, ela continua sendo
utilizada em diversas situa\c{c}\~oes em astrof\'{\i}sica e cosmologia.
\par
A emerg\^encia da teoria da relatividade restrita no in\'{\i}cio do s\'eculo XX levou \`a substitui\c{c}\~ao da mec\^anica newtoniana pela mec\^anica relativista.
Essa \'ultima, por sua vez, empregava como grupo de simetria fundamental o grupo de Lorentz, em vez do grupo de Galileu utilizado na mec\^anica newtoniana.
Ao mesmo tempo, em \'{\i}ntima rela\c{c}\~ao com o uso do grupo de Lorentz como estrutura matem\'atica fundamental, a mec\^anica relativista estabelecia que h\'a uma velocidade limite na natureza, a velocidade da luz $c$. No entanto, para velocidades muito inferiores \`a da luz, os resultados da mec\^anica relativista
s\~ao praticamente indistingu\'{\i}veis dos resultados da mec\^anica newtoniana. Isto faz com que o uso de mec\^anica relativista, na pr\'atica, ocorra apenas em algumas situa\c{c}\~oes, a maior parte delas obtidas em sofisticados laborat\'orios e aceleradores de part\'{\i}culas.
\par
A teoria da relatividade geral substituiu a gravita\c{c}\~ao newtoniana, sendo a moderna teoria da gravita\c{c}\~ao. Al\'em de incorporar conceitos da mec\^anica relativista, como a velocidade da luz como velocidade limite, a teoria da relatividade geral substitui a id\'eia de for\c{c}a gravitacional pela de curvatura do
espa\c{c}o-tempo. O grupo fundamental passa a ser o grupo de difeomorfismo oriundo da geometria diferencial, e que atua em variedades geom\'etricas. Mas, como ocorre no caso da mec\^anica relativista, a teoria gravitacional newtoniana \'e obtida da teoria da relatividade geral no limite em que as velocidades
s\~ao pequenas comparadas com a velocidade da luz e o campo gravitacional \'e fraco \footnote{O que significa um campo gravitacional {\it fraco}, conceito que requer um valor de refer\^encia, ser\'a definido mais tarde.}. Na maior parte dos casos, incluindo sistemas astron\^omicos como gal\'axias e aglomerados
de gal\'axias, a gravita\c{c}\~ao newtoniana pode ser usada sem maiores problemas. J\`a em menores escalas, sistemas estelares como an\~as brancas tamb\'em s\~ao bem descritas pela teoria newtoniana, mas objetos estelares super compactos, como estrelas de n\^eutrons, requerem o uso da teoria da relatividade geral. 
\par
Os estudos em cosmologia, que compreendem as maiores escalas conhecidas, revelam aspectos curiosos do uso ou da Relatividade Geral ou da teoria newtoniana, segundo o dom\'{\i}nio de aplica\c{c}\~ao de cada uma destas teorias. Quando a Relatividade Geral foi formulada com a estrutura que conhecemos hoje, em 1915, as informa\c{c}\~oes que se tinha sobre o que chamamos de {\it nosso universo} eram muito limitadas: sequer existia a no\c{c}\~ao da gal\'axia; os estudos cosmol\'ogicos eram incipientes, para n\~ao dizer inexistentes. A complexidade da nova teoria, seu car\'ater altamente n\~ao linear, for\c{c}ava a busca de solu\c{c}\~oes que exibiam altas
simetrias. O interesse evidente pelo estudo de estrelas (objetos sobre os quais j\'a se tinha muitas informa\c{c}\~oes), determinou a busca de solu\c{c}\~oes est\'aticas com simetria esf\'erica. Focalizou-se posteriormente em solu\c{c}\~oes que poderiam representar o universo, suposto inicialmente homog\^eneo, isotr\'opico e est\'atico, configura\c{c}\~ao que permitia se encontrar solu\c{c}\~oes exatas, mas que logo se revelaram inst\'aveis. Um pouco mais tarde, solu\c{c}\~oes din\^amicas, representando o que hoje n\'os denominamos universo homog\^eneo e isotr\'opico em expans\~ao, foram determinadas por Friedmann e Lema\^{\i}tre \cite{fried,lemaitre}.
\par 
A partir deste momento, e das descoberta que o universo \'e formado por gal\'axias, e que estas gal\'axias est\~ao se afastando umas das outras, caracterizando a expans\~ao c\'osmica \cite{lemaitre,hubble}, deu-se in\'{\i}cio aos estudos mais rigorosos de cosmologia. Tudo isto foi feito dentro do contexto da teoria da Relatividade 
Geral. Apenas na d\'ecada de 30, quando os estudos do universo feitos no contexto relativista j\'a adquiriam bases mais s\'olidas, foi que se tentou
construir uma cosmologia no contexto da teoria newtoniana: a teoria gravitacional mais antiga e mais simples encontrou sua aplica\c{c}\~ao em cosmologia depois que a teoria gravitacional mais recente e complexa tinha se apoderado deste campo de estudo \cite{milne1,milne2}.
\par
No entanto, a teoria newtoniana encontrou logo um espa\c{c}o de aplica\c{c}\~ao na cosmologia. Na descri\c{c}\~ao do universo que conhecemos hoje, que se expande e se esfria com o passar do tempo, o universo teria passado por quatro fases: a fase primordial, ainda sob estudos, e que talvez requeira a compreens\~ao de efeitos qu\^anticos em gravita\c{c}\~ao; a fase radiativa, onde o universo \'e dominado por um g\'as de f\'otons e de part\'{\i}culas relativistas; a fase material, na qual as estruturas c\'osmicas como as gal\'axias se formam, onde o universo \'e dominado por um fluido de mat\'eria sem press\~ao; e a fase atual, onde o universo parece ser dominado por um fluido ex\'otico, dito energia escura, que conduz \`a expans\~ao acelerada do universo. A teoria da Relatividade Geral tem uma caracter\'istica fundamental que a torna imprescind\'{\i}vel no estudo destas fases: a press\~ao do fluido desempenha um papel ativo, sendo tamb\'em fonte do campo gravitacional, algo que n\~ao tem equivalente na teoria newtoniana.
\par
No entanto, na fase material, onde a press\~ao \'e suposta nula, a teoria newtoniana poderia ser igualmente aplicada. Como esta \'e a fase de forma\c{c}\~ao das estruturas, o que exige complexos estudos perturbativos, a teoria newtoniana encontra ali uma aplica\c{c}\~ao conveniente, dada \`a sua simplicidade t\'ecnica e conceitual. Ali\'as, os modernos estudos de forma\c{c}\~ao de estruturas em um universo em expans\~ao utilizando simula\c{c}\~oes num\'ericas, requerem o uso da teoria newtoniana, mesmo que isto implique limita\c{c}\~oes ao se tentar introduzir componentes em que a press\~ao desempenha um papel mais importante, como \'e o caso da energia escura.
\par 
\'E poss\'{\i}vel estender os estudos newtonianos para as situa\c{c}\~oes onde a press\~ao desempenha um papel gravitacional ativo? Tenta-se responder a esta pergunta desde os anos 50. Se isto for poss\'{\i}vel, ter\'{\i}amos uma importante ferramenta em m\~aos, com impactos profundos nos estudos de forma\c{c}\~ao de estrutura no universo (incluindo as simula\c{c}\~oes num\'ericas), na determina\c{c}\~ao dos observ\'aveis em cosmologia e gravita\c{c}\~ao (que na teoria newtoniana possuem um sentido mais direto), e da consequente compara\c{c}\~ao da observa\c{c}\~ao com a teoria, quest\~ao central hoje nos estudos de fen\^omenos gravitacionais. Denominaremos tais poss\'{\i}veis extens\~oes da teoria newtoniana de {\it teorias neo-newtonianas} \cite{neo1,neo2,neo3}. Descrever
a busca de uma forma\c{c}\~ao consistente de uma teoria neo-newtoniana \'e o objeto deste texto. Sugerimos tamb\'em a leitura da refer\^encia \cite{rbef} sobre
o mesmo assunto.
\par
No que se segue, revisaremos primeiro a formula\c{c}\~ao newtoniana usual (pr\'oxima se\c{c}\~ao), a formula\c{c}\~ao relativista (se\c{c}\~ao 3),
ambas no contexto cosmol\'ogico. Na se\c{c}\~ao 4 discutiremos como podemos tentar incorporar a press\~ao como fonte do campo gravitacional, o que caracterizaria a teoria neo-newtoniana, discutindo suas aplica\c{c}\~ao \`a cosmologia e ao estudo de estrelas nas se\c{c}\~oes 5 e 6, respectivamente. Na se\c{c}\~ao 7, apresentamos nossas conclus\~oes.

\section{As equa\c{c}\~oes do fluido newtoniano em presen\c{c}a de um campo gravitacional}

A descri\c{c}\~ao de sistemas auto-gravitantes na teoria newtoniana se faz mais adequadamente usando uma descri\c{c}\~ao de fluidos, pelo menos para os prop\'ositos que temos em mente, a cosmologia e objetos estelares. Neste caso, o conjunto equa\c{c}\~oes \'e,
\begin{eqnarray}
\label{e1}
\frac{\partial\rho}{\partial t} + \nabla\cdot(\rho\vec v) &=& 0, \\
\label{e2}
\frac{\partial\vec v}{\partial t} + \vec v\cdot \nabla \vec v &=& - \frac{\nabla p}{\rho} - \nabla\phi,\\
\label{e3}
\nabla^2\phi &=& 4\pi G\rho.
\end{eqnarray}
Nestas equa\c{c}\~oes, $\rho$ representa a densidade do fluido, $p$ \'e a press\~ao correspondente, $\vec v$ \'e o campo de velocidade, $\phi$ \'e o potencial gravitacional. A equa\c{c}\~ao (\ref{e1}), denominada como {\it equa\c{c}\~ao da continuidade}, expressa a conserva\c{c}\~ao da mat\'eria. A equa\c{c}\~ao
(\ref{e2}), denominada {\it equa\c{c}\~ao de Euler}, nada mais \'e que a segunda lei de Newton re-expressa em termos das vari\'aveis do fluido, sendo que o lado
direito corresponde ao balan\c{c}o das for\c{c}as \`as quais o fluido est\'a submetido, neste caso o gradiente da press\~ao e a for\c{c}a gravitacional. A equa\c{c}\~ao (\ref{e3}), denominada {\it equa\c{c}\~ao de Poisson}, \'e a reformula\c{c}\~ao diferencial da lei do inverso do quadrado da dist\^ancia do campo gravitacional. Observe-se que apenas a mat\'eria \'e fonte do campo gravitacional, algo \'obvio no contexto newtoniano, mas menos evidente no contexto relativista.
\par
Consideremos agora a aplica\c{c}\~ao destas equa\c{c}\~oes \`a cosmologia. Neste caso, primeiramente tentamos incorporar os elementos b\'asicos do universo observado: ele \'e homog\^eneo e isotr\'opico em grandes escalas, e est\'a em expans\~ao. A homogeneidade e isotropia podem ser incorporadas
\`as equa\c{c}\~oes (\ref{e1},\ref{e2},\ref{e3}) supondo que a densidade e a press\~ao s\~ao fun\c{c}\~oes puramente do tempo: $\rho = \rho(t)$ e $p = p(t)$.
Por outro lado, a expans\~ao de um universo homog\^eneo e isotr\'opico pode ser descrita pela lei de Hubble, tal que,
\begin{eqnarray}
\vec v = H\vec r,
\end{eqnarray}
onde $\vec v$ \'e o campo de velocidade dos objetos que comp\~oem o universo, medido por um dado observador, e $\vec r$ \'e a dist\^ancia daqueles
objetos a esse observador. Devido \`a isotropria e homogeneidade, podemos supor que $H$, denominado fator de Hubble, depende unicamente do
tempo, $H = H(t)$. \'E conveniente escrever o fator de Hubble em termos de uma fun\c{c}\~ao $a = a(t)$, conhecida como {\it fator de escala}, tal que,
\begin{eqnarray}
H = \frac{\dot a}{a}.
\end{eqnarray}
\par
Inserindo estas defini\c{c}\~oes nas equa\c{c}\~oes (\ref{e1},\ref{e2},\ref{e3}), obtemos as seguintes equa\c{c}\~oes:
\begin{eqnarray}
\dot\rho + 3 \frac{\dot a}{a}\rho &=& 0,\\
\frac{\ddot a}{a} &=& - \frac{4\pi G}{3}\rho,\\
\nabla\phi &=& \frac{4\pi G}{3}\rho\vec r.
\end{eqnarray}
Observe-se que a press\~ao n\~ao se faz mais presente, o que \'e natural neste contexto, pois devido \`a homogeneidade e isotropia o gradiente de press\~ao \'e nulo.
A solu\c{c}\~ao dessas equa\c{c}\~oes \'e direta:
\begin{eqnarray}
a = a_0 t^{2/3}, \quad \rho = \rho_0\biggr(\frac{a}{a_0}\biggl)^3 = \rho_0 t^{-2}.
\end{eqnarray}
\par
\'E poss\'{\i}vel estudar perturbativamente essa configura\c{c}\~ao. Tal estudo \'e fundamental para a an\'alise do processo de forma\c{c}\~ao de estruturas (gal\'axias, aglomerados de gal\'axias, etc.) em um universo em expans\~ao. Isto \'e feito introduzindo pequenas flutua\c{c}\~oes em torno das solu\c{c}\~oes encontradas anteriormente:
\begin{eqnarray}
\tilde\rho = \rho + \delta\rho, \quad \tilde{\vec v} = \vec v + \delta\vec v, \quad \tilde\phi = \phi + \delta\phi, 
\end{eqnarray}
onde $\delta\rho$, $\delta p$, $\delta\vec v$ e $\delta\phi$ representam as pequenas flutua\c{c}\~oes em torno das solu\c{c}\~oes encontradas anteriormente.
Um c\'alculo relativamente longo, mas padr\~ao, retendo unicamente os termos lineares nas quantidades perturbadas, leva \`a equa\c{c}\~ao para
o {\it contraste na densidade} $\delta$, definido como,
\begin{eqnarray}
\delta = \frac{\delta\rho}{\rho},
\end{eqnarray}
tal que,
\begin{eqnarray}
\label{ep1}
\ddot\delta + 2 \frac{\dot a}{a}\dot\delta + \biggr\{k^2\frac{c^2_s}{a^2} - 4\pi G\rho\biggl\}\delta = 0.
\end{eqnarray}
Nessa express\~ao, $c_s^2 = \frac{\partial p}{\partial \rho}$ \'e a velocidade do som, e $k$ \'e o n\'umero de onda associado \`a decomposi\c{c}\~ao de Fourier, $f(t,\vec x) = f(t) e^{i\vec k\cdot\vec x}$.
\par
Note-se que a equa\c{c}\~ao (\ref{ep1}) tem a forma da equa\c{c}\~ao de um oscilador harm\^onico amortecido (o amortecimento ocorrendo gra\c{c}as
\`a expans\~ao do universo), onde dois efeitos competitivos definem o crescimento do contraste da densidade: a velocidade do som do fluido, que gera oscila\c{c}\~oes no comportamento de $\delta$ e a atra\c{c}\~ao gravitacional, que tende a induzir a condensa\c{c}\~ao das perturba\c{c}\~oes.
\par
Uma outra aplica\c{c}\~ao importante das equa\c{c}\~oes (\ref{e1},\ref{e2},\ref{e3}) se refere ao equil\'{\i}brio estelar. Neste caso, consideramos uma configura\c{c}\~ao est\'atica e radial, o que implica $\vec v = 0$ e que todas as fun\c{c}\~oes restantes ($\rho$, $p$ e $\phi$) dependem unicamente da coordenada radial. Assim, o sistema acima se reduz a ao seguinte conjunto de equa\c{c}\~oes:
\begin{eqnarray}
\frac{dp}{dr} &=& - \rho\frac{d\phi}{dr},\\
\frac{d\phi}{dr} &=& \frac{4\pi G}{r^2}\int_0^r \rho(r'){r'}^2\,dr'.
\end{eqnarray}
Estas equa\c{c}\~oes podem ser re-escritas como,
\begin{eqnarray}
\label{ee1}
\frac{dp}{dr} &=& - \frac{G\rho\,m(r)}{r^2},\\
\label{ee2}
m(r) &=& 4\pi\int_0^r \rho(r'){r'}^2\,dr'.
\end{eqnarray}
\par
Em geral, as estrelas podem ser modelizadas por fluidos barotr\'opicos, definidos como sendo aqueles em que a press\~ao depende apenas da densidade da mat\'eria, $p = p(\rho)$. Em particular, uma depend\^encia do tipo de lei de pot\^encia, do tipo $p = K\rho^n$, onde $K$ e $n$ s\~ao constantes, se revela bastante conveniente. As estrelas mais comuns, representadas no diagrama de Hertzprung-Russel, podem ser descritas, grosso modo, por estas express\~oes.
A equa\c{c}\~ao (\ref{ee1}) re-escrita usando os fluidos barotr\'opicos sob a forma de lei de pot\^encia, e feita sem dimens\~ao, \'e denominada de {\it equa\c{c}\~ao de Lane-Emden}.

\section{A teoria da Relatividade Geral}

Existem duas principais diferen\c{c}as da teoria da Relatividade Geral em rela\c{c}\~ao \`a sua antecessora, a gravita\c{c}\~ao newtoniana. Em primeiro lugar, o princ\'{\i}pio relativista de uma velocidade limite na natureza \'e incorporado. Em segundo lugar, a no\c{c}\~ao de for\c{c}a gravitacional \'e substitu\'{\i}da pela de curvatura do espa\c{c}o-tempo. A introdu\c{c}\~ao da velocidade limite torna a nova teoria gravitacional compat\'{\i}vel com a teoria da relatividade restrita. A geometriza\c{c}\~ao da intera\c{c}\~ao gravitacional permite introduzir na nova teoria o princ\'{\i}pio de equival\^encia, que estabelece na sua
forma fraca, que todos os corpos reagem da mesma forma ao campo gravitacional, independentemente  de sua massa. Assim, todos corpos seguem geod\'esicas no espa\c{c}o-tempo, suposto curvo e com estrutura geom\'etrica riemanianna. Desvios das trajet\'orias geod\'esticas ocorrem apenas quando os corpos s\~ao submetidos a for\c{c}as n\~ao-gravitacionais.
\par
A geometria da teoria da Relatividade Geral \'e definida no espa\c{c}o-tempo quadridimensional. A dist\^ancia infinitesimal entre dois pontos nesta geometria
\'e fornecida pela m\'etrica,
\begin{eqnarray}
ds^2 = g_{\mu\nu}dx^\mu\,dx^\nu,
\end{eqnarray}
onde adotamos a conven\c{c}\~ao da soma, segundo a qual \'{\i}ndices repetidos sup\~oem uma somat\'oria, e $g_{\mu\nu}$ s\~ao os coeficientes m\'etricos que definem localmente a geometria do espa\c{c}-tempo a quatro dimens\~oes.
\par
As equa\c{c}\~oes de Relatividade Geral s\~ao equa\c{c}\~oes tensoriais, onde o lado esquerdo est\'a relacionado \`a geometria do espa\c{c}o-tempo, e o lado direito descreve a distribui\c{c}\~ao de mat\'eria e energia.
Essas equa\c{c}\~oes t\^em a forma,
\begin{eqnarray}
\label{rg1}
R_{\mu\nu} - \frac{1}{2}g_{\mu\nu} R &=& \frac{8\pi G}{c^2} T_{\mu\nu}, \\
\label{rg2}
{T^{\mu\nu}}_{;\mu} &=& 0.
\end{eqnarray}
Na equa\c{c}\~ao (\ref{rg1}), $R_{\mu\nu}$ \'e o tensor de Ricci, definido por,
\begin{eqnarray}
R_{\mu\nu} = \partial_{\rho}\Gamma^\rho_{\mu\nu} - \partial_{\nu}\Gamma^\rho_{\mu\rho} + \Gamma^\rho_{\mu\nu}\Gamma^\sigma_{\rho\sigma} -
\Gamma^\rho_{\mu\sigma}\Gamma^\sigma_{\rho\nu},
\end{eqnarray}
sendo $\Gamma^\rho_{\mu\nu}$ a conex\~ao christoffeliana, definida por,
\begin{eqnarray}
\Gamma^\rho_{\mu\nu} = \frac{1}{2}g^{\rho\sigma}(\partial_\mu g_{\sigma\nu} + \partial_\nu g_{\sigma\nu} - \partial_\sigma g_{\mu\nu}).
\end{eqnarray}
O escalar de Ricci \'e definido como $R = g^{\rho\sigma}R_{\rho\sigma}$. Por sua vez, o tensor de momento-energia $T_{\mu\nu}$ depende do tipo
de mat\'eria ou campo que estamos considerando. Para um fluido perfeito com quadri-velocidade $u_\mu$, ele tem a forma,
\begin{eqnarray}
T_{\mu\nu} = \biggr(\rho + \frac{p}{c^2}\biggl)u_\mu u_\nu - p g_{\mu\nu}.
\end{eqnarray}
A equa\c{c}\~ao (\ref{rg2}) \'e consequ\^encia das chamadas identidades de Bianchi, que estabelece que a quadri-diverg\^encia do lado esquerdo da
equa\c{c}\~ao (\ref{rg1}) \'e identicamente zero.
\par
Na descri\c{c}\~ao cosmol\'ogica, usa-se o fato que o universo \'e, em grande escala, homog\^eneo e isotr\'opico. Assim, a m\'etrica tem a seguinte forma,
\begin{eqnarray}
ds^2 = c^2 dt^2 - a(t)^2(dx^2 + dy^2 + dz^2),
\end{eqnarray}
onde assumimos que o a se\c{c}\~ao espacial tri-dimensional \'e plana, e $a(t)$ \'e o fator de escala.
Neste caso,
as equa\c{c}\~oes (\ref{rg1},\ref{rg2}) se reduzem a,
\begin{eqnarray}
\label{erg1}
\biggr(\frac{\dot a}{a}\biggl)^2 &=& 8\pi G \rho,\\
\label{erg2}
2\frac{\ddot a}{a} + \biggr(\frac{\dot a}{a}\biggl)^2 &=& - 8\pi G\frac{p}{c^2},\\
\label{erg3}
\dot\rho + 3\frac{\dot a}{a}\biggr(\rho +\frac{ p}{c^2}\biggl) &=& 0.
\end{eqnarray}
Devido \`as identidades de Bianchi, apenas duas destas equa\c{c}\~oes s\~ao independentes.
Observe que a press\~ao desempenha um papel determinante no comportamento de $a(t)$ ao contr\'ario do que ocorre na cosmologia newtoniana descrita na se\c{c}\~ao anterior.
\par
As solu\c{c}\~oes de (\ref{erg1},\ref{erg2},\ref{erg3}) dependem da equa\c{c}\~ao de estado do fluido. Para uma depend\^encia linear da press\~ao com a
densidade, $p = \omega\rho c^2$, $\omega$ constante, essas equa\c{c}\~oes podem ser facilmente resolvidas, implicando em,
\begin{eqnarray}
a = a_0 t^\frac{2}{3(1 + \omega)}.
\end{eqnarray}
Quando $\omega = - 1$ (que corresponde a chamada equa\c{c}\~ao de estado do v\'acuo qu\^antico), o fator de escala cresce exponencialmente.
Apenas quando a press\~ao \'e nula, $\omega = 0$, a solu\c{c}\~ao relativista coincide com a solu\c{c}\~ao newtoniana. No entanto, para o caso de um universo dominado por um fluido radiativo, caracterizado por $\omega = 1/3$, o fator de escala evolui como $a \propto t^{1/2}$, o que difere do caso
newtoniano. Para um universo dominado por energia escura, para o qual $\omega < - 1/3$ a diferen\c{c}a \'e ainda mais pronunciada: o universo expande
aceleradamente, algo imposs\'{\i}vel de se obter no caso newtoniano, devido \`a natureza puramente atrativa da gravita\c{c}\~ao. Lembramos que as observa\c{c}\~oes indicam que o Universo deve ser atualmente dominado por energia escura.
\par
As dificuldades conceituais e t\'ecnicas da teoria da Relatividade Geral se tornam consider\'aveis quando se procura estudar perturba\c{c}\~oes em torno 
do modelo cosmol\'ogico descrito acima. Um dos motivos \'e a alta n\~ao-linearidade das equa\c{c}\~oes da Relatividade Geral. Outro motivo \'e que essas
equa\c{c}\~oes s\~ao invariantes por difeomorfismo, o que implica invari\^ancia por transforma\c{c}\~oes gerais de coordenadas. Este \'ultimo fato gera dificuldades em identificar quais s\~ao as perturba\c{c}\~oes f\'{\i}sicas e quais s\~ao os efeitos da escolha de um sistema de coordenadas. Estas dificuldades
geraram diversos formalismos para abordar o problema perturbativo em cosmologia, e identificar os observ\'aveis f\'{\i}sicos.
No entanto, o problema se simplifica consideravelmente quando a press\~ao \'e nula. Neste caso, o contraste na densidade se comporta
como,
\begin{eqnarray}
\ddot\delta + 2\frac{\dot a}{a}\dot\delta - 4\pi G(1 + \omega)(1 + 3\omega)\rho\delta = 0.
\end{eqnarray}
Esta equa\c{c}\~ao perturbada se assemelha com a newtoniana correspondente mostrada na se\c{c}\~ao precedente, em duas situa\c{c}\~oes; quando a velocidade do som \'e nula (o que implica novamente press\~ao nula), e neste caso as equa\c{c}\~oes coincidem; quando o n\'umero de onda $k$ \'e nulo, o que implica perturba\c{c}\~oes em grandes escalas. Neste \'ultimo caso, no entanto, as solu\c{c}\~oes s\~ao diferentes do caso newtoniano correspondente, devido ao comportamento do fator de escala; 
a correspond\^encia \'e completa unicamente no caso de press\~ao nula, novamente.
\par
O caso do equil\'{\i}brio de uma estrela \'e bem mais complexo na teoria da Relatividade Geral.
A equa\c{c}\~ao do equil\'{\i}brio hidrodin\^amico estelar l\^e-se, neste caso, como \cite{wein},
\begin{eqnarray}
\frac{dP}{dr} = - \frac{G}{r^2} \biggr(\rho + \frac{p}{c^2}\biggl)\frac{ m(r) + 4\pi r^3 \frac{p}{c^2}}{\biggr(1 - \frac{2Gm(r)}{r c^2}\biggl)}
\end{eqnarray}
Esta \'e equa\c{c}\~ao {\it TOV}, acr\^onimo para Tolman-Oppenheimer-Volkoff, os primeiros autores a estudar o problema do equil\'{\i}brio estelar no contexto da relatividade geral. A equa\c{c}\~ao TOV se reduz \`a equa\c{c}\~ao newtoniana correspondente apenas quando se toma o limite $c \rightarrow \infty$, o que implica desconsiderar a exist\^encia de uma velocidade da luz na natureza, condi\c{c}\~ao imposta pelo princ\'{\i}pio relativista. De uma forma geral, as condi\c{c}\~oes para o equil\'{\i}brio estelar em Relatividade Geral s\~ao bem diferentes das condi\c{c}\~oes newtonianas correspondentes. Essas diferen\c{c}as, no entanto, s\~ao desprez\'{\i}veis quando a press\~ao n\~ao \'e muito importante, comparada com a densidade de mat\'eria, o que ocorre para boa
parte das estrelas que formam a sequ\^encia principal no diagrama de Hertzsprung-Russel. No entanto, para objetos compactos, como as estrelas de n\^eutrons, os resultados relativistas s\~ao diferentes dos newtonianos, tanto qualitativa quanto quantitativamente. 
\par
Neste momento podemos definir o que entendemos por campo fraco (quando a teoria newtoniana pode ser usada) e campo forte (quando a teoria relativista
deve for\c{c}ocamente ser usada). Para isto, definimos a grandeza sem dimens\~ao $k = GM/(R c^2)$, que nada mais \'e que o potencial gravitacional dividido pela velocidade da luz ao quadrado. Esta quantidade fornece a raz\~ao entre o efeito gravitacional e o efeito relativista. Quando $k << 1$ temos
o campo fraco; quando $k$ \'e da ordem da unidade os efeitos relativistas s\~ao consider\'aveis e estamos no regime de campo forte.

\section{As teorias neo-Newtonianas}

A teoria da Relatividade Geral descrita na se\c{c}\~ao precedente tem como base o princ\'{\i}pio da equival\^encia. O princ\'{\i}pio da equival\^encia estabelece
que todos os corpos reagem ao campo gravitacional da mesma forma, independentemente de sua massa. Uma das consequ\^encias diretas disto \'e que, localmente, o campo gravitacional \'e indistingu\'{\i}vel de um referencial acelerado. A forma de se introduzir a universalidade do princ\'{\i}pio da equival\^encia \'e geometrizar a intera\c{c}\~ao gravitacional, fazendo com que o campo gravitacional seja apenas um efeito da curvatura do espa\c{c}o-tempo a quatro dimens\~oes. Neste caso, todos os corpos sob a\c{c}\~ao apenas da gravita\c{c}\~ao seguir\~ao geod\'esicas nesta geometria curva. Consequentemente, todos os corpos reagem igualmente \`a gravita\c{c}\~ao e o princ\'{\i}pio da equival\^encia est\'a automaticamente incorporado.
\par
O princ\'{\i}pio da equival\^encia \'e baseado na igualdade entre a massa inercial, cujo conceito est\'a relacionado \`a segunda lei da mec\^anica newtoniana,
e a massa gravitacional, definida pela lei da gravita\c{c}\~ao newtoniana. Podemos, no entanto, definir tr\^es tipos de massa: a inercial, definida acima, a gravitacional passiva e a gravitacional ativa. A massa inercial indica como um corpo reage a uma for\c{c}a arbitr\'aria conforme a famosa lei de for\c ca $\vec F=m\vec a$; a massa gravitacional passiva determina como um corpo reage ao campo gravitacional; e a massa gravitacional ativa indica como um corpo cria o campo gravitacional. O princ\'{\i}pio da equival\^encia, na sua forma fraca, implica a igualdade entre a massa inercial e a massa gravitacional passiva. No entanto, existem outras formula\c{c}\~oes do princ\'{\i}pio da equival\^encia, podendo implicar a igualdade dos tr\^es tipos de massa.
\par
Nosso objetivo agora \'e tentar criar uma formula\c{c}\~ao da teoria newtoniana tal que efeitos t\'{\i}picos da Relatividade Geral sejam incorporados em um contexto espa\c{c}o-temporal newtoniano (tempo como par\^ametro externo universal, espa\c{c}o tridimensional euclidiano). Uma das consequ\^encias da teoria Relatividade, enfatizado na se\c{c}\~ao anterior, \'e o papel gravitacional desempenhado pela press\~ao, isto quando se usa um fluido ou campo como fonte do campo gravitacional. Uma tentativa neste sentido seria identificar a massa inercial e a massa gravitacional passiva como sendo dada por $\rho + p$, ao passo que a massa gravitacional ativa como sendo $\rho + 3p$. Tais identifica\c{c}\~oes parecem arbitr\'arias, no entanto a nova express\~ao para a massa gravitacional ativa est\'a intimamente relacionada com a no\c{c}\~ao de {\it condi\c{c}\~ao de energia forte}, que diz, no contexto da teoria da Relatividade Geral, quando uma configura\c{c}\~ao gravitacional (considerando a press\~ao) possui efeito atrativo ou repulsivo.
\par
Segundo a proposta de constru\c{c}\~ao de uma teoria neo-newtoniana, incorporando efeitos relativistas \`a teoria newtoniana, descrita acima, as novas
equa\c{c}\~oes da continuidade, de Euler e de Poisson teriam a seguinte forma:
\begin{eqnarray}
\label{nn1-a}
\frac{\partial\rho}{\partial t} + \nabla\cdot[(\rho + p)\vec v] &=& 0,\\
\label{nn1-b}
\frac{\partial\vec v}{\partial t} + \vec v\cdot\nabla \vec v &=& - \frac{\nabla p}{\rho + p} - \nabla\phi,\\
\label{nn1-c}
\nabla^2\phi &=& 4\pi G(\rho + 3p).
\end{eqnarray}
Observe-se que a corrente de mat\'eria e a segunda lei de Newton incorporam os novos conceitos de massa inercial e massa gravitacional passiva,
ao passo que a nova equa\c{c}\~ao de Poisson utiliza o novo conceito de massa in\'ercia ativa. Por simplicidade, nas expresss\~oes acima utilizamos um sistema de coordenadas onde $c = 1$.
\par
Uma outra possibilidade de se construir uma teoria neo-newtoniana da gravita\c{c}\~ao consiste em utilizar argumentos termodin\^amicos ao se rescrever a
equa\c{c}\~ao da continuidade. De fato, a identfiica\c{c}\~ao descrita acima para a massa inercial \'e incompleta: na equa\c{c}\~ao da continuidade, o termo com derivada temporal (associado \`a massa contida em um volume $V$) continua tendo a forma anterior. Isto foi feito para guardar contato com o limite
da teoria da Relatividade Geral na aproxima\c{c}\~ao de campo fraco e baixas velocidades, quando a teoria deve se reduzir \`a newtoniana no limite de ordem zero, com corre\c{c}\~oes em primeira ordem.
No entanto, podemos modificar a equa\c{c}\~ao da continuidade ao evocar a primeira lei da termodin\^amica e o papel que a press\~ao nela exerce: a press\~ao est\'a relacionada ao trabalho realizado quando o volume do sistema se expande. A expans\~ao do volume por sua vez est\'a relacionado ao divergente
do campo de velocidade.
\par
Baseado nas considera\c{c}\~oes acima, podemos propor o novo conjunto de equa\c{c}\~oes como sendo o seguinte:
\begin{eqnarray}
\label{nn2-a}
\frac{\partial\rho}{\partial t} + \nabla\cdot(\rho\vec v) + p\nabla\cdot\vec v &=& 0,\\
\label{nn2-b}
\frac{\partial\vec v}{\partial t} + \vec v\cdot\nabla \vec v &=& - \frac{\nabla p}{\rho + p} - \nabla\phi,\\
\label{nn2-c}
\nabla^2\phi &=& 4\pi G(\rho + 3p).
\end{eqnarray}
\par
As equa\c{c}\~oes (\ref{nn1-a},\ref{nn1-b},\ref{nn1-c}) definir\~ao o que chamaremos daqui por diante de {\it teoria neo-newtoniana tipo I} (NNI), enquanto
as equa\c{c}\~oes (\ref{nn2-a},\ref{nn2-b},\ref{nn2-c}) definir\~ao a {\it teoria neo-newtoniana tipo II} (NNII). Nas pr\'oximas se\c{c}\~oes analisaremos as consequ\^encias destas equa\c{c}\~oes para a cosmologia e a condi\c{c}\~ao de equil\'{\i}brio estelar.

\section{Teorias neo-newtonianas em Cosmologia}

Como j\'a comentado anteriormente, uma abordagem newtoniana para a evolu\c c\~ao do cosmos utiliza conceitos da mec\^anica dos fluidos. Nesse caso, adotando o princ\'ipio cosmol\'ogico e a lei de Hubble, define-se o campo de velocidades deste fluido que comp\~oe  o universo como sendo $\vec{v}= H \vec{r}$. Al\'em disto, a press\~ao e a densidade dependem unicamente do tempo. Quando utilizamos, neste contexto, as equa\c{c}\~oes newtonianas usuais, (\ref{e1},\ref{e2},{\ref{e3}), obtemos um comportamento que independe da press\~ao. Isto era esperado, pois em um universo homog\^eneo e isotr\'opico o gradiente de press\~ao \'e nulo, e consequentemente as solu\c{c}\~oes obtidas s\~ao as mesmas para qualquer tipo de express\~ao que consideramos para a press\~ao. No entanto, as equa\c{c}\~oes da Relatividade Geral dependem crucialmente da press\~ao, mesmo sob as hip\'oteses de homogeneidade e isotropia. No jarg\~ao usual, a press\~ao {\it gravita}.
\par
As modifica\c{c}\~oes introduzidas no \^ambito da teoria newtoniana por suas extens\~oes ditas neo-newtonianas modificam substancialmente essa situa\c{c}\~ao: as equa\c c\~oes (\ref{nn1-a},\ref{nn1-b},\ref{nn1-c}) e (\ref{nn2-a},\ref{nn2-b},\ref{nn2-c}) fornecem exatamente o mesmo tipo de evolu\c{c}\~ao para a din\^amica de fundo do universo, quando se considera as hip\'oteses de homogeneidade e isotropia. Basicamente, obtemos as leis de Friedmann para o fator de escala $a(t)$, descritas pelas equa\c{c}\~oes (\ref{erg1},\ref{erg2},\ref{erg3}). A real diferen\c ca entre estes conjuntos de equa\c{c}\~oes surge ao se estudar o comportamento das pequenas perturba\c c\~oes associadas a densidade de mat\'eria $\delta \rho$. Como nesse caso existem gradientes das perturba\c c\~oes da press\~ao, percebe-se que (\ref{nn1-a}) e (\ref{nn2-a}) podem conduzir a resultados distintos. Disso, como veremos a seguir, chega-se a conclus\~ao que NNII \'e o sistema ideal para a cosmologia.
\par
A introdu\c{c}\~ao de pequenas perturba\c{c}\~oes em torno da solu\c{c}\~ao cosmol\'ogica no contexto do conjunto de equa\c{c}\~oes (\ref{nn1-a},\ref{nn1-b},\ref{nn1-c}), fornece (ap\'os a lineariza\c{c}\~ao descrita na se\c{c}\~ao 2 e da decomposi\c{c}\~ao em modos de Fourier) a seguinte equa\c{c}\~ao para a evolu\c{c}\~ao do contraste na densidade\cite{neo3}:
\begin{eqnarray}
\ddot\delta + 2\frac{\dot a}{a}\dot\delta + \biggr\{\frac{c_s^2}{a^2}k^2 - 4\pi G(1 + \omega)(1 + 3\omega)\rho\biggl\}\delta = - \frac{\ddot a}{a}\omega k^2\delta - \frac{\dot a}{a}\omega \vec k\cdot\nabla\dot\delta. 
\end{eqnarray}
Ao deduzir esta equa\c{c}\~ao consideramos uma equa\c{c}\~ao de estado do tipo $p = \omega\rho c^2$, com $\omega$ constante.
Esta \'e uma equa\c{c}\~ao bastante distinta da que deduzimos anteriormente, sobretudo devido aos termos que aparecem no seu lado direito da igualdade.
No entanto, quando a press\~ao \'e nula, ou quando as escalas s\~ao suficientemente grandes tais que o n\'umero de onda pode ser desprezado, ent\~ao reobtemos a mesma equa\c{c}\~ao que na teoria da Relatividade Geral, sob condi\c{c}\~oes similares.
\par
No caso da teoria NNII, a equa\c{c}\~ao perturbada correspondente que descreve a evolu\c{c}\~ao do contraste na densidade, assume a seguinte forma:
\begin{eqnarray}
\ddot\delta + 2\frac{\dot a}{a}\dot\delta + \biggr\{\frac{c_s^2}{a^2}k^2 - 4\pi G(1 + \omega)(1 + 3\omega)\rho\biggl\}\delta = 0. 
\end{eqnarray}
Esta ainda n\~ao \'e a equa\c{c}\~ao relativista (cuja forma, \'e preciso lembrar, depende do formalismo utilizado). Em geral, a equival\^encia pode n\~ao ser completa \cite{ribamar}. No entanto, neste novo caso as diferen\c{c}as s\~ao muito menos not\'aveis. Em alguns casos, por\'em, as diferen\c{c}as s\~ao muito pouco importantes, e isto n\~ao apenas para perturba\c{c}\~oes em grandes
escalas: perturba\c{c}\~oes em pequena escalas tamb\'em podem ser enfocadas coerentemente na teoria NNII.
\par
A \'ultima afirma\c{c}\~ao acima foi demonstrando no artigo \cite{ds}, onde se analisou o caso de um universo homog\^eneo e isotr\'opico preenchido por
um fluido viscoso. Utilizando tal componente dissipativa, tenta-se dar aos fluidos c\'osmicos um car\'ater mais realista. A viscosidade foi descrita no \^ambito do formalismo de Eckart. No caso relativista a press\~ao viscosa assume a forma,
\begin{eqnarray}
p = - \xi_0\rho^\nu uˆ\mu_{;\mu} = - 3\frac{\dot a}{a}\xi_0\rho^\nu,
\end{eqnarray}
onde $\xi_0$ e $\nu$ s\~ao constantes, e a hip\'otese de um universo homog\^eno e isotr\'opico foi utilizada.
No contexto newtoniano, esta express\~ao \'e dada por,
\begin{eqnarray}
p = - \xi_0\rho^\nu \nabla\cdot\vec v = - 3\frac{\dot a}{a}\xi_0\rho^\nu,
\end{eqnarray}
\par
Quando se usa a teoria NNII, obt\'em-se, ap\'os um c\'adulo longo, mas direto, a seguinte express\~ao para a evolu\c{c}\~ao no contraste na densidade:
\begin{eqnarray}
\ddot{\delta}+\left(2H-3 H \omega-\frac{\dot{\omega}}{1+\omega}\right)\dot{\delta}+\left[-4\pi G \rho \left(1+\omega\right)-6 H^2 w-3\dot{H}\omega-\frac{3H \dot{\omega}}{1+\omega}\right]\delta = \\ \nonumber
\frac{\nabla^2 \delta p}{a^2\rho} -3H\frac{\dot{\delta p}}{\rho}+\frac{\delta p}{\rho}\left[12\pi G \rho(1+\omega)-15 H^2-9H^2 \omega+\frac{3H\dot{\omega}}{1+\omega}-3\dot{H}\right].
\end{eqnarray}
Nesta equa\c{c}\~ao, $H = \frac{\dot a}{a}$, $\omega = p/\rho$, e $\delta p$ \'e a perturba\c{c}\~ao do termo da press\~ao viscosa.
\par
A equa\c{c}\~ao relativista correspondente no cal\'bre s\'ncrono n\~ao possui uma forma simples em termos de uma equa\c{c}\~ao \'unica. Ao contr\'ario, as perturba\c{c}\~oes
obedecem ao seguinte sistema de equa\c{c}\~oes:
\begin{equation}\label{rcont}
\dot{\delta}= -(1+\omega)\left(\frac{\theta}{a}-3\dot{\phi}\right)+3H\omega\delta-3H\frac{\delta p}{\rho};
\end{equation}
\begin{equation}\label{reuler}
\dot{\theta}= -H(1-3\omega)\theta-\frac{\dot{\omega}}{1+\omega}\theta+\frac{k^2}{a\left(1+\omega\right)}\frac{\delta p}{\rho}+\frac{k^2}{a}\phi,
\end{equation}
onde $\theta=i k^j v_j$ \'e a diverg\^encia do velocidade perturbada do fluido e $\phi$ est\'a relacionado \`as perturba\c{c}\~oes na m\'etrica.
\par
Apesar da estrutura aparentemente muito diferente do caso neo-newtoniano e do caso relativista, quando se considera perturba\c{c}\~oes que se encontram na faixa das observa\c{c}\~oes dispon\'{\i}veis hoje (aproximadamente entre dezenas e centenas de megaparsecs), os resultados s\~ao muito similares, como
mostra, a t\'{\i}tulo puramente exemplificativo, a figura $1$.

\begin{figure}\label{fig1}	
\begin{center}
\includegraphics[width=0.4\textwidth]{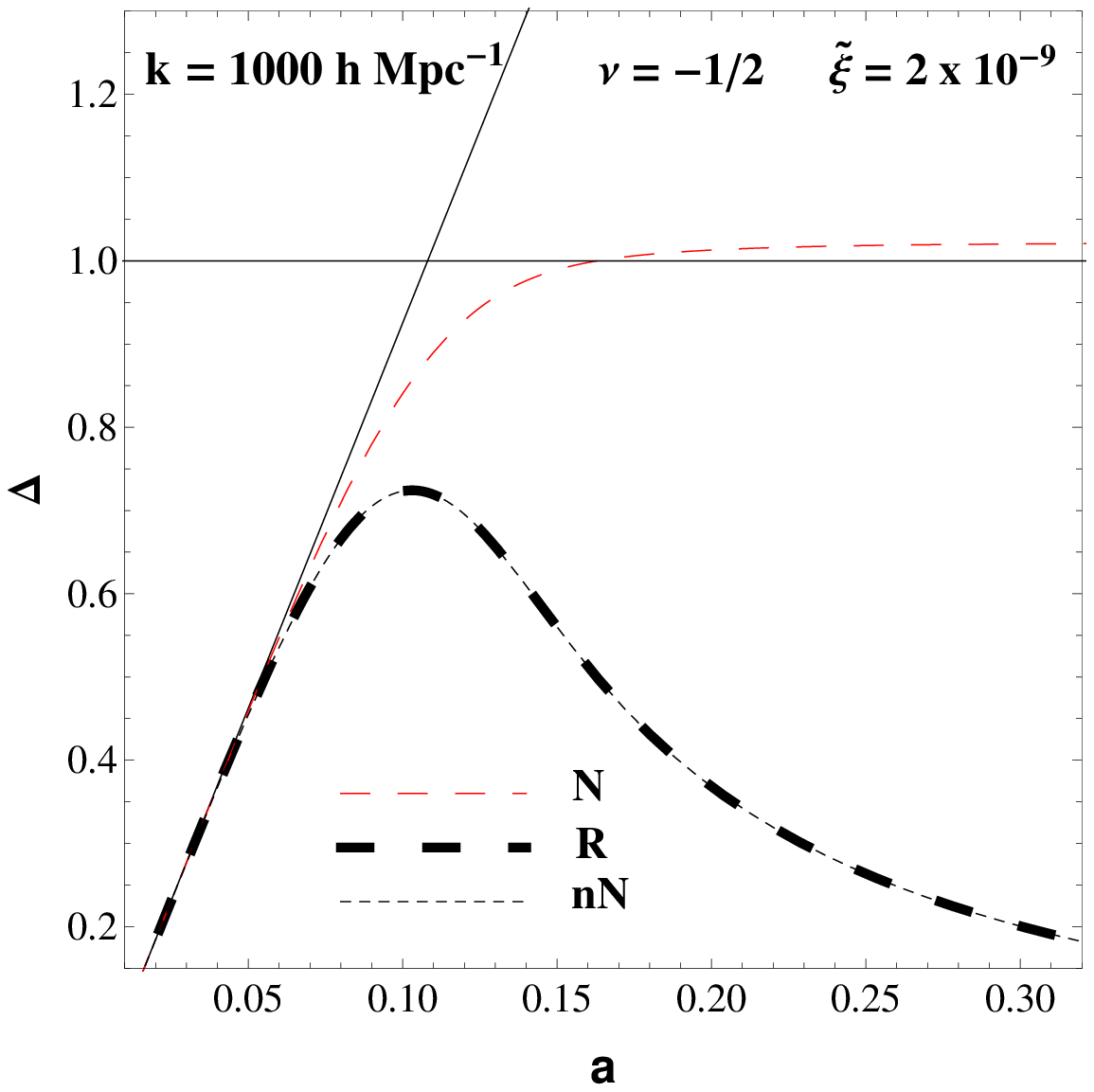}
\includegraphics[width=0.4\textwidth]{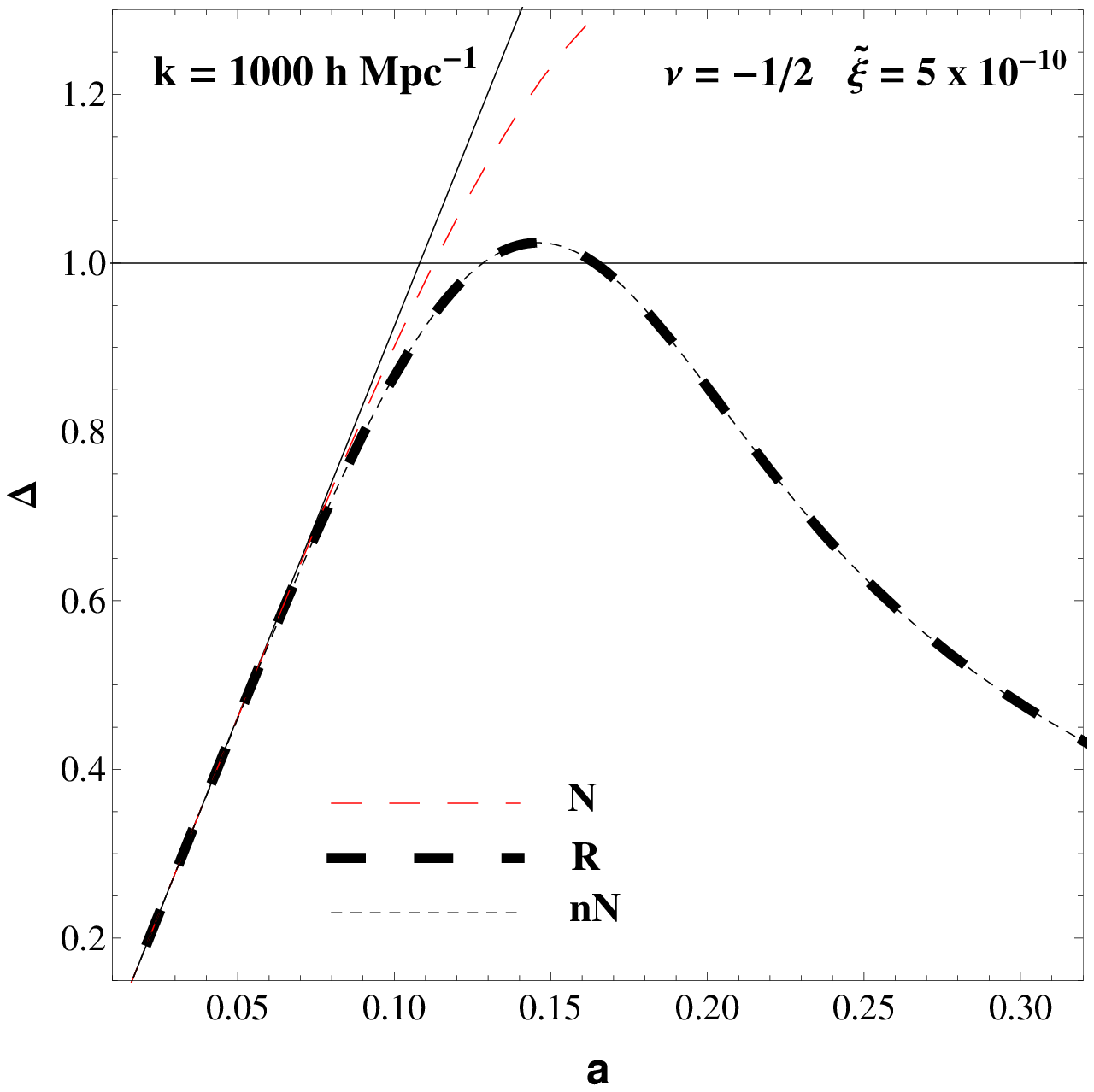}
\includegraphics[width=0.4\textwidth]{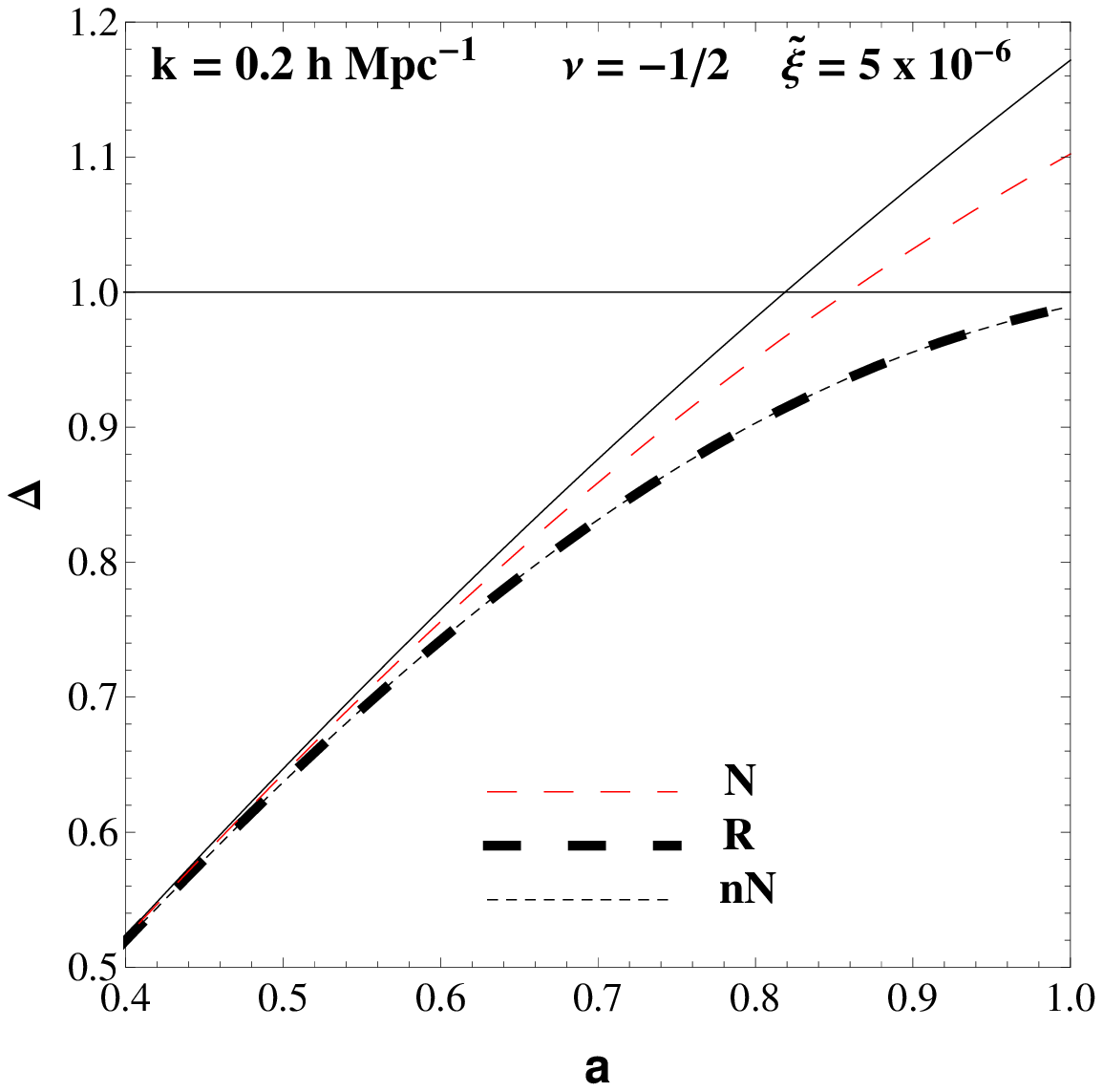}
\includegraphics[width=0.4\textwidth]{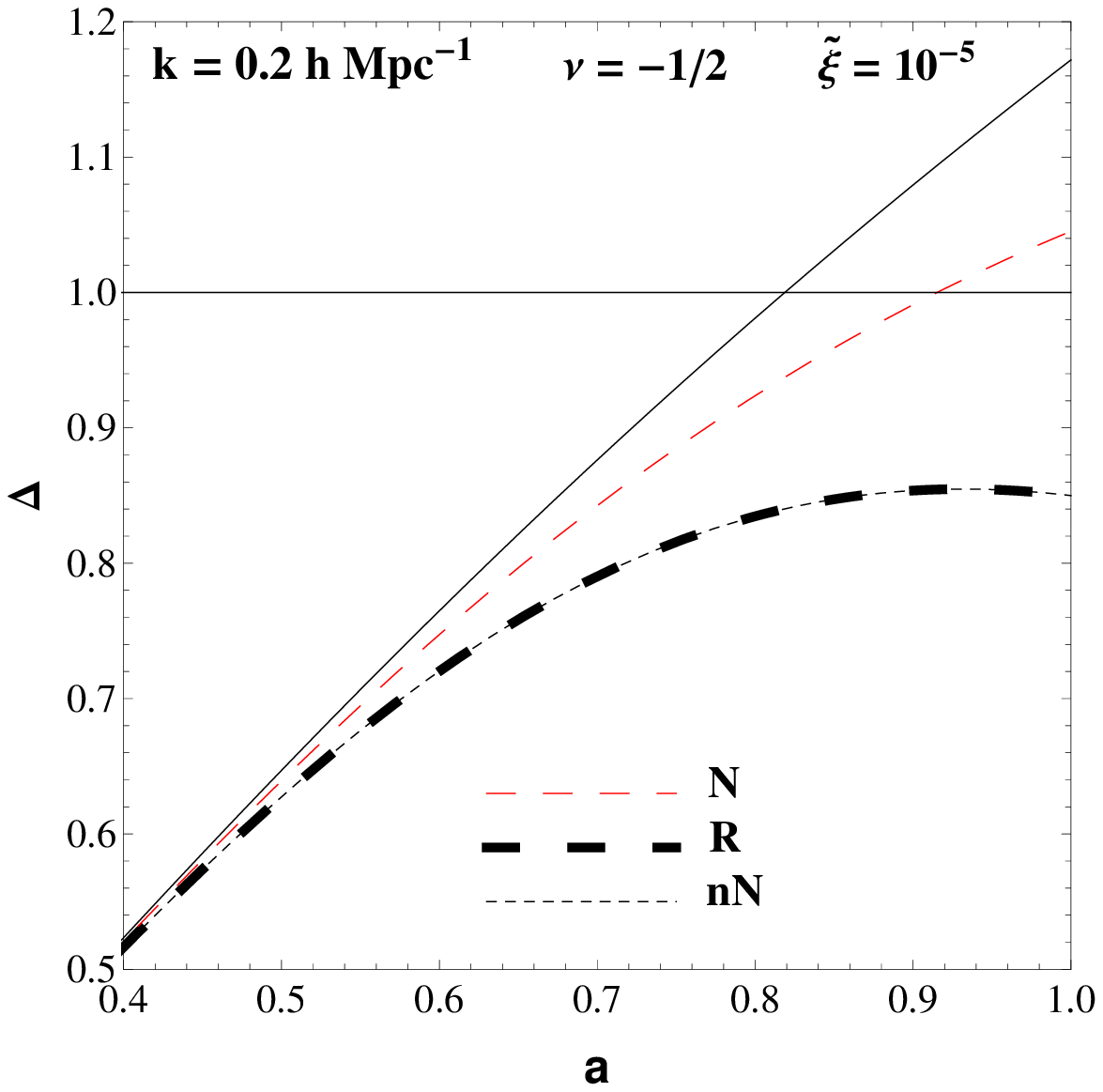}
\caption{Crescimento das perturba\c{c}\~oes correspondentes a gal\'axis an\~as, $k=1000 h {\rm Mpc}^{-1}$ (pain\'eis superiores) e escalas gal\'aticas, $k=0.2 h {\rm Mpc}^{-1}$ (pain\'eis inferiores) assumindo $\nu=-1/2$ na teoria de Eckart. Linhas s\'olidas correspondente ao modelo cosmol\'ogico padr\~ao ($\Lambda$CDM), ao passo que a linha vermelha pontilhada refere-se \`a teoria newtoniana usual para o caso viscoso, a linha azul para a teoria neo-newtoniana e a linha pontilhada preta ao
caso relativista viscoso. As linhas
horizontais delimitam o regime n\~ao linear.}
\end{center}
\end{figure}

O fato que, para escalas de interesse observacionais os resultados da teoria NNII s\~ao essencialmente id\^enticos aos relativistas, abre perspectivas muito interessantes, sobretudo no que diz respeito ao estudo de simula\c{c}\~oes num\'ericas. Estes estudos procuram determinar a forma\c{c}\~ao das estruturas locais (gal\'axias por exemplo), e requerem uma formula\c{c}\~ao newtoniana. No momento, se emprega o formalismo newtoniano usual. Mas, existe possibilidade de se ter, a partir dos resultados acima, um formalismo neo-newtoniano consistente.

\section{Teorias neo-newtonianas e o equil\'ibrio estelar}

O problema de equil\'{\i}brio estelar \'e outro campo importante de aplica\c{c}\~ao de teorias gravitacionais. O equil\'{\i}brio de uma estrela \'e mantido pelo balan\c{c}o do gradiente da press\~ao no seu interior, provocado pelas rea\c{c}\~oes nucleares que geram a energia emitida pela estrela, e a atra\c{c}\~ao
gravitacional, que tende a colapsar a estrela. As diversas fases da vida de uma estrela, representadas em sua maioria no diagrama de Hertzsprung-Russel, revelam os diversos mecanismos de {\it queima} da fus\~ao nuclear no interior das estrelas. Estas fus\~oes s\~ao respons\'aveis pela gera\c{c}\~ao de elementos qu\'{\i}micos do H\'elio at\'e o ferro. A partir do ferro, os processos de explos\~ao estelar (supernovas) \'e que sintetizaram estes elementos mais pesados. O deut\'erio, por sua vez, \'e sintetizado no universo primordial, em seus primeiros minutos de exist\^encia.
\par
Para as estrelas ordin\'arias, o valor absoluto da press\~ao n\~ao \'e compar\'avel \`a densidade de energia, mesmo que o seu gradiente possa ser muito grande. No entanto, a vida de uma estrela pode conduzir, no seu fim, a objetos compactos. O primeiro seriam as an\~as brancas, estrela {\it fria} que n\~ao gera mais energia e cuja configura\c{c}\~ao de equil\'{\i}brio \'e determinada pela degeneresc\^encia qu\^antica eletr\^onica. Apesar de fazer uso de conceitos 
tipicamente qu\^anticos para a compreens\~ao de seu estado de equil\'{\i}brio, a an\~a branca ainda pode ser descrita pela gravita\c{c}\~ao newtoniana.
\par
O segundo objeto compacto, com raio e densidade muito maiores que as an\~as brancas, seriam as estrelas de neutr\^ons, onde a atra\c{c}\~ao gravitacional \'e compensada pela degeresc\^encia qu\^antica dos n\^eutrons. Neste caso, os efeitos gravitacionais relativistas j\'a se tornam consider\'aveis: estrelas de
n\^eutrons possuem massa de algumas massas solares, comprimidas em um volume de raio de alguns quil\^ometros. Isto implica densidades da ordem 
de,
\begin{eqnarray}
\rho = \frac{M}{V} \sim \frac{10^{33}}{10^{18}}\frac{g}{cm^3} \sim 10^{15} \frac{g}{cm^3},
\end{eqnarray}
densidade esta que \'e t\'{\i}pica da mat\'eria nuclear.
\par
Em geral, para definir a import\^ancia dos efeitos relativistas se usa o indicador dimensional definido na se\c{c}\~ao 3, $k = \frac{GM}{R c^2}$. Quando um objeto, com raio $R$ e massa $M$ possui um $k \stackrel{>}{\sim} 1$, os efeitos relativistas n\~ao podem ser ignorados. No caso das estrelas de neutr\^ons, $k \sim 0,1$. Nesta situa\c{c}\~ao, o uso da teoria da Relatividade Geral para descrever as estrelas de neutr\^ons n\~ao pode ser evitado.
\par
A equa\c{c}\~ao que descreve o equil\'{\i}brio estela na teoria gravitacional newtoniana \'e a de Lane-Emden, mostrada ao fim da se\c{c}\~ao 2. A sua correspondente relativista \'e a equa\c{c}\~ao TOV mostrada na se\c{c}\~ao 3. No caso das teorias neo-newtonianas, podemos tamb\'em deduzir uma equa\c{c}\~ao de equil\'{\i}brio estelar utilizando as hip\'oteses de estaticidade ($\vec v = 0$) e de depend\^encia apenas da coordenada radial para a densidade, press\~ao e potencial gravitacional. Seguindo os mesmos passos que conduziram \`a equa\c{c}\~ao de Lane-Emden, obtemos a seguinte equa\c{c}\~ao para o equil\'{\i}brio estelar para as duas teorias neo-newtonianas descritas na se\c{c}\~ao 4:
\begin{eqnarray}
\frac{dp}{dr} &=& - \frac{G m(r)}{r^2}(\rho + p),\\
\frac{dm(r)}{dr} &=& 4 \pi Gr^2 (\rho + 3p). 
\end{eqnarray}
Essas equa\c{c}\~oes diferem tanto da sua correspondente newtoniana quanto da equa\c{c}\~ao TOV da Relatividade Geral. No entanto, elas t\^em uma
estrutura mais pr\'oxima da equa\c{c}\~ao relativista, com a press\~ao tendo um papel mais relevante que no caso puramente newtoniano. 

\begin{figure}[!t]
\begin{center}
\includegraphics[width=0.5\textwidth]{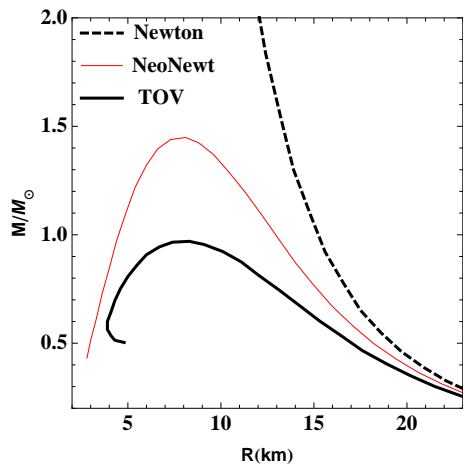}
\end{center}
\caption{Diagrama massa-raio para as estrelas de neutr\^ons utilizando a teoria newtoniana, a Relatividade Geral (TOV) e a teoria neo-newtoniana. O raio
\'e dado em quil\^ometros e a massa em unidades de massa solar, $M_\odot$.}
\end{figure}

Na figura 2 utilizamos um modelo bem simples para a configura\c c\~ ao estelar onde apenas neutr\^ons formam um g\'as de Fermi degenerado no interior estelar. Mostramos o diagrama massa-raio para as estrelas de neutr\^ons, para os tr\^es casos: o newtoniano, o relativista e o neo-newtoniano \cite{adriano}. A equa\c{c}\~ao newtoniana prev\^e a possibilidade de forma\c{c}\~ao de objetos indefinidamente compactos, em contradi\c{c}\~ao com a observa\c{c}\~ao, que prev\^e um limite neste diagrama, mostrado pelo m\'aximo no diagrama massa-raio; as teorias neo-newtonianas reproduzem o resultado qualitativo relativista, mas prev\^em a possibilidade de objetos com maior densidade que no caso relativista.
\par
Na an\'alise neo-newtoniana h\'a uma dificuldade que merece ser citada. Temos tr\^es defini\c{c}\~oes de densidade: $\rho$, $\rho + p$ e $\rho + 3p$.
Qual delas deveremos utilizar no diagrama massa-raio? Esta \'e uma quest\~ao cuja resposta \'e pouco clara. No caso da figura 2 utilizamos a defini\c c\~ao usual, que coincide com a newtoniana. O emprego das outras possibilidades preservaria qualitativamente os resultados descritos acima, mas modificaria
os valores absolutos mostrados na figura 2. 

\section{Observa\c c\~oes Finais}

Neste ano de 2015 se comemoram os 100 anos da teoria da Relatividade Geral, a moderna teoria da gravita\c{c}\~ao apresentada na refer\^encia \cite{eins}
em 1915. A teoria da Relatividade Geral visava corrigir dois aspectos incompletos da teoria gravitacional newtoniana: a aus\^encia da no\c{c}\~ao de uma velocidade limite na natureza, a velocidade da luz, necess\'aria devido \`a universalidade da teoria da Relatividade Restrita; incorporar de forma consistente
o princ\'{\i}pio da equival\^encia, que estabelece a igualdade entre as massas inercial e gravitacional, conforme indicado pela experi\^encia. A teoria da Relatividade Geral logrou tamb\'em explicar alguns fen\^omenos gravitacionais, que do ponto de vista newtoniano apareciam como anomalias n\~ao recebendo nenhuma explica\c{c}\~ao plaus\'{\i}vel, como o avan\c{c}o do peri\'elio de Merc\'urio.
A teoria da Relatividade Geral recebeu formid\'aveis confirma\c{c}\~oes observacionais e experimentais, como descrito na refer\^encia \cite{will}.
\par
No entanto, a teoria newtoniana continua sendo ainda empregada na maior parte das situa\c{c}\~oes. Em primeiro lugar, porque a teoria da Relatividade Geral
conduz na maior parte dos casos a corre\c{c}\~oes pequenas - e, frequentemente, negligenci\'aveis - \`a teoria newtoniana. Em segundo lugar, porque o
arcabou\c{c}o matem\'atico e conceitual da teoria newtoniana \'e sensivelmente mais simples; a teoria newtoniana tamb\'em se acorda mais facilmente
\`a intui\c{c}\~ao comum.
\par
Por outro lado, existem situa\c{c}\~oes onde \'e imposs\'{\i}vel ignorar a teoria da Relatividade Geral. Dois exemplos s\~ao a cosmologia, na maior parte
das fases da evolu\c{c}\~ao c\'osmica, e os objetos astrof\'{\i}sicos compactos, como estrelas de neutr\^ons. No caso da cosmologia, existe uma coincid\^encia entre as predi\c{c}\~oes newtonianas e relativistas para a fase material, quando a press\~ao \'e nula, mas nas fases primoridal, radiativa e acelerada do universo, a descri\c{c}\~ao newtoniana \'e inteiramente inadequada, principalmente porque a teoria newtoniana n\~ao \'e capaz de incorporar efeitos
da press\~ao em um universo homog\^eneo e isotr\'opico. No caso de objetos compactos, e especificamente no caso das estrelas de neutr\^ons, a teoria
newtoniana \'e incapaz de prever a rela\c{c}\~ao massa-raio observada. Devemos ainda acrescentar que a teoria newtoniana n\~ao \'e capaz de predizer
a exist\^encia de buracos negros, objetos possivelmente existentes em sistemas astrof\'{\i}sicos e gal\'aticos, que se revelam estruturas essencialmente
relativistas.
\par
No entanto, a teoria newtoniana continua a ter um papel a desempenhar devido \`as j\'a mencionadas simplicidades matem\'aticas e conceituais, e seria
desej\'avel estender a formula\c{c}\~ao newtoniana de forma a incorporar efeitos t\'{\i}picos da Relatividade Geral, isto sem comprometer os aspectos
matem\'aticos e conceituais da teoria newtoniana usual. Tais extens\~oes caracterizam as teorias neo-newtonianas discutidas neste texto. A chave para
criar estas extens\~oes se baseia em modificar a maneira como a massa \'e introduzida na teoria newtoniana, definindo uma nova massa inercial e uma
nova massa gravitacional que incorporem os efeitos da press\~ao.
\par
Discutimos duas possibilidades de se proceder a estas extens\~oes, levando \`as teorias neo-newtonianas tipo I e II apresentadas anteriormente. A teoria
tipo II revelou-se mais promissora, sobretudo devido \`as suas aplica\c c\~oes \`a cosmologia: al\'em de reproduzir a evolu\c{c}\~ao do universo n\~ao perturbado em todas suas fases, ela fornece resultados perturbativos coerentes com os obtidos na teoria da Relatividade Geral, pelo menos para
as escalas de interesse observacional. No entanto, apesar de reproduzir qualitativamente o diagrama massa-raio das estrelas de neutr\^ons, do ponto
de vista quantitativo, algumas descrep\^ancias aparecem. Isto revela que a cria\c{c}\~ao de uma teoria neo-newtoniana que incorpore o essencial dos
efeitos relativistas em um contexto matem\'afico e conceitual newtoniano \'e um projeto ainda em andamento. Por sinal, al\'em das propostas
apresentadas aqui, outras propostas surgem na literatura \cite{noh}. 
\par
O fato \'e que a constru\c{c}\~ao de uma teoria neo-newtoniana seguindo o plano delineado acima teria fortes impactos sobre os estudos de simula\c{c}\~ao num\'erica, que visam a reproduzir as estruturas n\~ao lineares observadas no universo, assim como o estudo de buracos negros an\'alogos \cite{ines}.
Esperamos que este programa de pesquisa conduza a resultados de alta relev\^ancia cient\'{\i}fica em um futuro pr\'oximo.
\newline
\vspace{0.3cm}
\newline
{\bf Agradecimentos}: JCF e HESV agradecem ao CNPq e \`a FAPES por apoio financeiro parcia. JCF agradece aos organizadores do {\bf Encontro
de F\'{\i}sica da Amaz\^onia Caribenha} pela calorosa hospitalidade em Boa Vista, Roraima, durante este evento.

\end{document}